%% file: main.tex
\documentclass[letterpaper]{easychair}
\usepackage{packages}
\input{newcommands}

\title{Realizing Implicit Computational Complexity%
	\thanks{\href{https://spots.augusta.edu/caubert/research/statycc/}{This research} is supported by the \href{https://face-foundation.org/higher-education/thomas-jefferson-fund/}{Th. Jefferson Fund} of the Embassy of France in the United States and the \href{https://face-foundation.org/}{FACE Foundation}, and has benefited from the research meeting \href{https://www.dagstuhl.de/de/programm/kalender/evhp/?semnr=21453}{21453 \enquote{Static Analyses of Program Flows: Types and Certificate for Complexity}} in Schloss Dagstuhl. Th.\ Rubiano and Th.\ Seiller are supported by the Île-de-France region through the DIM RFSI project \enquote{CoHOp}.}
}

\titlerunning{Realizing Implicit Computational Complexity}

\author{
	Clément Aubert\inst{1} \and
	Thomas Rubiano\inst{2} \and 
	Neea Rusch\inst{1} \and
	Thomas Seiller\inst{2,3}
}

\institute{
	School of Computer and Cyber Sciences, Augusta University, USA
\and
	LIPN – UMR 7030 Université Sorbonne Paris Nord, France
\and
	CNRS, France
 }

\authorrunning{C. Aubert, Th. Rubiano, N. Rusch and Th. Seiller}

\begin{document}

\maketitle


\paragraph{Originalities in Implicit Computational Complexity.}
Automatic performance analysis and optimization is a critical for systems with 
resource constraints.
The field of Implicit Computational Complexity (ICC)~\cite{DalLago2012a} pioneers in embedding in the program itself a guarantee of its resource usage, using \eg bounded recursion~\cite{Bellantoni1992,Leivant1993} or type systems~\cite{Baillot2004,Lafont2004}. It drives better understanding of complexity classes, but also introduces original methods to develop resources-aware languages, static source code analyzers and optimizations techniques, often relying on informative and subtle type systems. Among the methods developed, the \emph{mwp-flow analysis}~\cite{Jones2009} certifies polynomial bounds on the size of the values manipulated by an imperative program, obtained by bounding the transitions between states instead of focusing on states in isolation, and is not concerned with termination or tight bounds on values. It introduces a new way of tracking dependencies between \enquote{chunks} of code by typing each statement with a matrix listing the way variables relate to each others. 

Having introduced such novel analysis techniques, and, as opposed to traditional complexity, by utilizing models that are generally expressive enough to write down actual algorithms~\cite[p.~11]{Moyen2017c}, ICC provides a conceivable pathway to automatable complexity analysis and optimization. However, the approaches have rarely materialized into concrete programming languages or program analyzers extending beyond toy languages, with a few exceptions~\cite{Avanzini2017,Hoffmann2012b}. Absence of realized results reduces ability to test the true power of these techniques, limits their application in general, and understanding their capabilities and potential expressivity remains underexplored. 

We present an ongoing effort to address this deficiency by applying the mwp-flow analysis, that tracks dependencies between variables, in three different ways, at different stages of maturation, in their temporal order. The first and third projects bend this typing discipline to gain finer-grained view on statements independence, to optimize loops by hoisting invariant~\cite{Moyen2017} and by splitting loops \enquote{horizontally} to parallelize them~\cite{Aubert2022f}. The second project refines, extends and implements the original analysis to obtain a fast, modular static analyzer~\cite{Aubert2022b}. All three projects aim at pushing the original type system to its limits, to assess how ICC can in practice lead to original, sometimes orthogonal, approaches. We also discuss our intent and motivations behind formalizing this analysis using Coq proof assistant~\cite{coqman}, in a spearheading endeavour toward formalizing complexity analysis.

\subparagraph{1. Loop Quasi-Invariant Chunk Detection.}
Loop peeling for hoisting (quasi-)invariants can be used to optimize generated code~\cite[p. 641]{Aho2006}, and is implemented \eg in LLVM as the \href{https://llvm.org/docs/Passes.html#licm-loop-invariant-code-motion}{\texttt{licm}} pass.
This work~\cite{Moyen2017} leverages an ICC-inspired dependency analysis to provide a transformation method to compilers. It enables detection of quasi-invariants of arbitrary degree in composed statements called \enquote{chunks}. It reuses the mwp's matricial system and typed data flows to generate dependency graphs, to compute an invariance degree for each statement or chunks of statements. It then finds the maximum (worst) dependency graph for loops, and recognizes whether an entire block is quasi-invariant. If this block is an inner loop, it can be hoisted, and the computational complexity of the overall program can be decreased.
A prototype analysis pass~\cite{lqicm} has been designed, proven correct and successfully implemented using a toy \texttt{C} parser, and as a prototype pass for the LLVM. This is the first known application of introducing ICC techniques in mainstream compilers.

\subparagraph{2. Improved and Implemented mwp-Analysis.}
In an ongoing development, we improved and implemented the mwp-bounds analysis~\cite{Jones2009}, which certifies that the values computed by an imperative program are bounded by polynomials in the program's input, represented in a matrix of typed flows, characterizing controls from one variable to another. While this flow analysis is elegant and sound, it is also computationally costly--it manipulates non-deterministically a potentially exponential number of matrices in the size of the program~\cite[2.3]{Aubert2022b}---and missed an opportunity to leverage its built-in compositionality. We addressed both issues by expanding the original flow calculus, and adjusting its internal machinery to enable tractable analysis~\cite{Aubert2022b}, and further extended the theory with analysis of function definitions and calls---including recursive ones, a feature not widely supported~\cite[p.~359]{Hainry2021}.  Our effort and theoretical development is realized in an open-source tool \texttt{pymwp}~\cite{pymwp}, capable of automatically analyzing complexity of programs written in a subset of the \texttt{C} programming language.

\subparagraph{3. Splitting Loops Horizontally to Improve Their Parallel Treatment.}
Our most recent effort is directed toward program optimization through loop parallelization. Using an ICC-inspired data flow-based variable dependency analysis, we can reproduce the \emph{tour de force} of detecting opportunities for loop fission that have been missed by other standard analyses~\cite{Moyen2017}. In particular, the dependency analysis allows optimizing loops by splitting them \enquote{horizontally}, \eg from \pr|for (int i = 1; i < 10; i++){a[i] = a[i-1] + i; b[i] = b[i-1] + i;}| to:

\begin{lstlisting}[language=C, backgroundcolor = \color{white}]
for (int i = 1; i < 10; i++){a[i] = a[i-1] + i;}
for (int i = 1; i < 10; i++){b[i] = b[i-1] + i;}
\end{lstlisting}
Our approach can process loop-carried dependencies~\cite[3.5.2]{chandra2000}--such as the one illustrated above--and optimizes \pr|while| loops~\cite[Sect. 5]{Aubert2022f}.---and, more generally, loops whose trip-count cannot be known at compilation time---that are completely ignored by OpenMP~\cite[3.2.2]{chandra2000}, and generally present great difficulty and often 
prevent optimization
~\cite[Sect. A]{Aubert2022f}.
Combined with OpenMP \pr|pragma| directives, this approach gives a speed-up \enquote{as good as} \texttt{AutoPar-Clava}---which \enquote{compare\textins{s} favorably with closely related auto-parallelization compilers}~\cite[p.~1]{Arabnejad2020}---when both are applicable that can be integrated in automatic parallelization pipelines~\cite[Sect. B]{Aubert2022f}.
Our benchmark, shared at \url{https://github.com/statycc/icc-fission}, substantiate experimentally those claims and provides further evidence.

\subparagraph{\dots and Pushing Even Further.}
From there, many other directions can be explored. Since ICC techniques tend to be designed for simpler program syntax, compiler intermediate representations present an ideal location and point of integration for performing analyses. Implementing the analysis in certified tools such as the CompCert compiler~\cite{Leroy2009} (or, more precisely, its static single assignment version~\cite{Barthe2014}) 
naturally necessitates certifying the complexity analysis, and we plan to pursue this effort using the Coq proof assistant~\cite{coqman}. The plasticity of both compilers and of the implemented analysis should facilitate porting our results to support further programming languages in addition to \texttt{C}.  As complexity analysis is 
difficult in Coq~\cite{Gueneau2019}, we believe a push 
would be welcome, and that ICC provides the necessary tools for it.

\newpage
\bibliographystyle{plainurl}
\bibliography{bib}
\end{document}

%% file: newcommands.tex
\theoremstyle{remark}
\AtBeginDocument{

}

\newcommand{\pr}{\lstinline[mathescape]}

\ebproofnewstyle{small}{
	separation = 1em, rule margin = .5ex,
}

\newcommand*{\eg}{e.g.\@\xspace}

\makeatletter
\newcommand*{\etc}{%
	\@ifnextchar{.}%
	{etc}%
	{etc.\@\xspace}%
}
\makeatother

\usepackage{soul}

\makeatletter
\renewcommand{\boxed}[1]{\text{\fboxsep=.2em\fbox{\m@th$\displaystyle#1$}}}
\makeatother



%% file: main.bbl
\begin{thebibliography}{10}

\bibitem{Aho2006}
Alfred~V. Aho, Monica~S. Lam, Ravi Sethi, and Jeffrey~D. Ullman.
\newblock {\em Compilers: Principles, Techniques, and Tools (2nd Edition)}.
\newblock Addison Wesley, August 2006.

\bibitem{Arabnejad2020}
Hamid Arabnejad, Jo{\~{a}}o Bispo, Jo{\~{a}}o M.~P. Cardoso, and Jorge~G.
  Barbosa.
\newblock Source-to-source compilation targeting openmp-based automatic
  parallelization of {C} applications.
\newblock {\em J.\ Supercomput.}, 76(9):6753--6785, Sep 2020.
\newblock \href {https://doi.org/10.1007/s11227-019-03109-9}
  {\path{doi:10.1007/s11227-019-03109-9}}.

\bibitem{lqicm}
Clément Aubert, Thomas Rubiano, Neea Rusch, and Thomas Seiller.
\newblock Lqicm on c toy parser.
\newblock URL: \url{https://github.com/statycc/LQICM_On_C_Toy_Parser}.

\bibitem{pymwp}
Clément Aubert, Thomas Rubiano, Neea Rusch, and Thomas Seiller.
\newblock {pymwp: MWP analysis in Python}.
\newblock URL: \url{https://github.com/statycc/pymwp/}.

\bibitem{Aubert2022f}
Clément Aubert, Thomas Rubiano, Neea Rusch, and Thomas Seiller.
\newblock {A Novel Loop Fission Technique Inspired by Implicit Computational
  Complexity}.
\newblock Submitted to \href{https://atva-conference.org/2022/}{ATVA 2022}, May
  2022.
\newblock URL: \url{https://hal.archives-ouvertes.fr/hal-03669387}.

\bibitem{Aubert2022b}
Clément Aubert, Thomas Rubiano, Neea Rusch, and Thomas Seiller.
\newblock mwp-analysis improvement and implementation: Realizing implicit
  computational complexity.
\newblock In Amy Felty, editor, {\em 7th International Conference on Formal
  Structures for Computation and Deduction (FSCD)}, LIPIcs. Schloss Dagstuhl,
  March 2022.
\newblock To appear.
\newblock URL: \url{https://hal.archives-ouvertes.fr/hal-03596285}.

\bibitem{Avanzini2017}
Martin Avanzini and Ugo~Dal Lago.
\newblock Automating sized-type inference for complexity analysis.
\newblock {\em Proc.\ ACM Program.\ Lang.}, 1({ICFP}):43:1--43:29, 2017.
\newblock \href {https://doi.org/10.1145/3110287} {\path{doi:10.1145/3110287}}.

\bibitem{Baillot2004}
Patrick Baillot and Kazushige Terui.
\newblock Light types for polynomial time computation in lambda-calculus.
\newblock In {\em LICS}, pages 266--275. IEEE Computer Society, 2004.
\newblock \href {https://doi.org/10.1109/LICS.2004.1319621}
  {\path{doi:10.1109/LICS.2004.1319621}}.

\bibitem{Barthe2014}
Gilles Barthe, Delphine Demange, and David Pichardie.
\newblock Formal verification of an {SSA}-based middle-end for compcert.
\newblock {\em ACM Trans.\ Program.\ Lang.\ Syst.}, 36(1):4:1--4:35, 2014.
\newblock \href {https://doi.org/10.1145/2579080} {\path{doi:10.1145/2579080}}.

\bibitem{Bellantoni1992}
Stephen~J. Bellantoni and Stephen~Arthur Cook.
\newblock A new recursion-theoretic characterization of the polytime functions
  (extended abstract).
\newblock In S.~Rao Kosaraju, Mike Fellows, Avi Wigderson, and John~A. Ellis,
  editors, {\em STOC}, pages 283--93. ACM, 1992.
\newblock \href {https://doi.org/10.1145/129712.129740}
  {\path{doi:10.1145/129712.129740}}.

\bibitem{chandra2000}
Rohit Chandra, Ramesh Menon, Leo Dagum, David Kohr, Dror Maydan, and Jeff
  McDonald.
\newblock {\em Parallel Programming in {OpenMP}}.
\newblock Morgan Kaufmann, Oxford, England, October 2000.

\bibitem{DalLago2012a}
Ugo Dal~Lago.
\newblock A short introduction to implicit computational complexity.
\newblock In Nick Bezhanishvili and Valentin Goranko, editors, {\em ESSLLI},
  volume 7388 of {\em LNCS}, pages 89--109. Springer, 2011.
\newblock \href {https://doi.org/10.1007/978-3-642-31485-8_3}
  {\path{doi:10.1007/978-3-642-31485-8_3}}.

\bibitem{Gueneau2019}
Arma{\"{e}}l Guéneau.
\newblock {\em Mechanized Verification of the Correctness and Asymptotic
  Complexity of Programs. (Vérification mécanisée de la correction et
  complexité asymptotique de programmes)}.
\newblock PhD thesis, Inria, Paris, France, 2019.
\newblock URL: \url{https://tel.archives-ouvertes.fr/tel-02437532}.

\bibitem{Hainry2021}
Emmanuel Hainry, Emmanuel Jeandel, Romain Péchoux, and Olivier Zeyen.
\newblock Complexityparser: An automatic tool for certifying poly-time
  complexity of {J}ava programs.
\newblock In Antonio Cerone and Peter~Csaba {\"{O}}lveczky, editors, {\em
  Theoretical Aspects of Computing - {ICTAC} 2021 - 18th International
  Colloquium, Virtual Event, Nur-Sultan, Kazakhstan, September 8-10, 2021,
  Proceedings}, volume 12819 of {\em LNCS}, pages 357--365. Springer, 2021.
\newblock \href {https://doi.org/10.1007/978-3-030-85315-0_20}
  {\path{doi:10.1007/978-3-030-85315-0_20}}.

\bibitem{Hoffmann2012b}
Jan Hoffmann, Klaus Aehlig, and Martin Hofmann.
\newblock Resource aware {ML}.
\newblock In P.~Madhusudan and Sanjit~A. Seshia, editors, {\em Computer Aided
  Verification - 24th International Conference, {CAV} 2012, Berkeley, CA, USA,
  July 7-13, 2012 Proceedings}, volume 7358 of {\em LNCS}, pages 781--786.
  Springer, 2012.
\newblock \href {https://doi.org/10.1007/978-3-642-31424-7_64}
  {\path{doi:10.1007/978-3-642-31424-7_64}}.

\bibitem{Jones2009}
Neil~D. Jones and Lars Kristiansen.
\newblock A flow calculus of \emph{mwp}-bounds for complexity analysis.
\newblock {\em ACM Trans.\ Comput.\ Log.}, 10(4):28:1--28:41, 2009.
\newblock \href {https://doi.org/10.1145/1555746.1555752}
  {\path{doi:10.1145/1555746.1555752}}.

\bibitem{Lafont2004}
Yves Lafont.
\newblock Soft linear logic and polynomial time.
\newblock {\em Theor.\ Comput.\ Sci.}, 318(1):163--180, 2004.
\newblock \href {https://doi.org/10.1016/j.tcs.2003.10.018}
  {\path{doi:10.1016/j.tcs.2003.10.018}}.

\bibitem{Leivant1993}
Daniel Leivant.
\newblock Stratified functional programs and computational complexity.
\newblock In Mary~S. Van~Deusen and Bernard Lang, editors, {\em Conference
  Record of the Twentieth Annual ACM SIGPLAN-SIGACT Symposium on Principles of
  Programming Languages}, pages 325--333. {ACM} Press, 1993.
\newblock \href {https://doi.org/10.1145/158511.158659}
  {\path{doi:10.1145/158511.158659}}.

\bibitem{Leroy2009}
Xavier Leroy.
\newblock Formal verification of a realistic compiler.
\newblock {\em Commun.\ ACM}, 52(7):107--115, 2009.
\newblock \href {https://doi.org/10.1145/1538788.1538814}
  {\path{doi:10.1145/1538788.1538814}}.

\bibitem{Moyen2017c}
Jean{-}Yves Moyen.
\newblock {\em Implicit Complexity in Theory and Practice}.
\newblock Habilitation thesis, University of Copenhagen, 2017.
\newblock URL: \url{https://lipn.univ-paris13.fr/~moyen/papiers/
  Habilitation_JY_Moyen.pdf}.

\bibitem{Moyen2017}
Jean{-}Yves Moyen, Thomas Rubiano, and Thomas Seiller.
\newblock Loop quasi-invariant chunk detection.
\newblock In Deepak D'Souza and K.~Narayan Kumar, editors, {\em Automated
  Technology for Verification and Analysis - 15th International Symposium,
  {ATVA} 2017, Pune, India, October 3-6, 2017, Proceedings}, volume 10482 of
  {\em LNCS}. Springer, 2017.
\newblock \href {https://doi.org/10.1007/978-3-319-68167-2_7}
  {\path{doi:10.1007/978-3-319-68167-2_7}}.

\bibitem{coqman}
The Coq~Development Team.
\newblock The coq proof assistant, version 8.7.0, October 2017.
\newblock \href {https://doi.org/10.5281/zenodo.1028037}
  {\path{doi:10.5281/zenodo.1028037}}.

\end{thebibliography}
